\documentclass[letterpaper]{article} 
\usepackage{aaai25}  
\usepackage{times}  
\usepackage{helvet}  
\usepackage{courier}  
\usepackage[hyphens]{url}  
\usepackage{graphicx} 
\usepackage{natbib}  
\usepackage{caption} 
\usepackage{algorithm}
\usepackage{algorithmic}
\usepackage{multirow}
\usepackage{multicol}
\usepackage{booktabs}
\usepackage{tabularx}
\usepackage{adjustbox}
\usepackage{amsfonts}
\usepackage{amsmath}
\usepackage{color}
\usepackage[T1]{fontenc}

\usepackage{newfloat}
\usepackage{listings}
\DeclareCaptionStyle{ruled}{labelfont=normalfont,labelsep=colon,strut=off} 
\floatstyle{ruled}
\newfloat{listing}{tb}{lst}{}
\floatname{listing}{Listing}
\title{CoDe: Communication Delay-Tolerant Multi-Agent Collaboration via Dual Alignment of Intent and Timeliness}
\author{
    Shoucheng Song\textsuperscript{\rm 1,2}, 
    Youfang Lin\textsuperscript{\rm 1,2},
    Sheng Han\textsuperscript{\rm 1,2},
    Chang Yao\textsuperscript{\rm 1,2},
    Hao Wu\textsuperscript{\rm 1,2},\\
    Shuo Wang\textsuperscript{\rm 1,2},
    Kai Lv\textsuperscript{\rm 1,2}\thanks{Corresponding Author: Kai Lv (lvkai@bjtu.edu.cn)}\\
}
\affiliations{
    \textsuperscript{\rm 1}School of Computer Science \& Technology, Beijing Jiaotong University, Beijing, China\\
    \textsuperscript{\rm 2}Beijing Key Laboratory of Traffic Data Mining and Embodied Intelligence, Beijing, China\\
    \{insis\_songsc, yflin, shhan, 22120453, 21241019, 19112029, lvkai\}@bjtu.edu.cn
}

\usepackage{bibentry}
\begin{document}

\maketitle

\begin{abstract}
Communication has been widely employed to enhance multi-agent collaboration. 
Previous research has typically assumed delay-free communication, a strong assumption that is challenging to meet in practice. 
However, real-world agents suffer from channel delays, receiving messages sent at different time points, termed {\it{Asynchronous Communication}}, leading to cognitive biases and breakdowns in collaboration.
This paper first defines two communication delay settings in MARL and emphasizes their harm to collaboration.
To handle the above delays, this paper proposes a novel framework \textbf{Co}mmunication \textbf{De}lay-tolerant Multi-Agent Collaboration (CoDe).
At first, CoDe learns an intent representation as messages through future action inference, reflecting the stable future behavioral trends of the agents.
Then, CoDe devises a dual alignment mechanism of intent and timeliness to strengthen the fusion process of asynchronous messages. 
In this way, agents can extract the long-term intent of others, even from delayed messages, and selectively utilize the most recent messages that are relevant to their intent.
Experimental results demonstrate that CoDe outperforms baseline algorithms in three MARL benchmarks without delay and exhibits robustness under fixed and time-varying delays.
\end{abstract}

%

\section{Introduction}
Communication is crucial for enhancing the collaborative capability of multi-agents, especially within distributed systems. Through internal communication, agents restricted by local observations can strengthen their understanding of teammates and the environment, leading to more cooperative behaviors \cite{foerster2016learning,das2019tarmac,ding2020learning,yuan2022multi,sun2024t2mac}. 
However, previous works frequently rely on ideal communication conditions, such as unlimited bandwidth, noise-free channels, and delay-free transmissions. These strong assumptions are typically hard to achieve in the real world.

Prior efforts have primarily focused on communication architectures to achieve sparse communication under bandwidth constraints, which involves gating mechanisms \cite{jiang2018learning, han2023model}, event-triggered mechanisms \cite{hu2020event}, and predefined rules \cite{zhang2019efficient,zhang2020succinct,sheng2022learning,yuan2022multi}. Moreover, several studies 
\cite{mitchell2020gaussian,xue2021mis,yuan2024communication, ding2024learning, yu2024robust} also consider the impact of channel noise and malicious attacks and make further enhancements. 

Nevertheless, only a few researchers have concentrated on the delay issue in MARL. DAMARL \cite{chen2020delay} addresses interaction delays between agents and the environment, assuming agents receive observations after a fixed time interval, without involving inter-agent communication. DACOM \cite{yuan2023dacom} centers on communication delays among agents, integrating waiting time into the decision process. It assumes the delay is shorter than a single decision interval, thus ignoring message asynchronicity issues. Actually, channel delays often extend across decision intervals in more general scenarios, meaning that agents receive and utilize delayed messages from varying time points. 

\begin{figure}[t]
\centering
\includegraphics[width=1.0\linewidth]{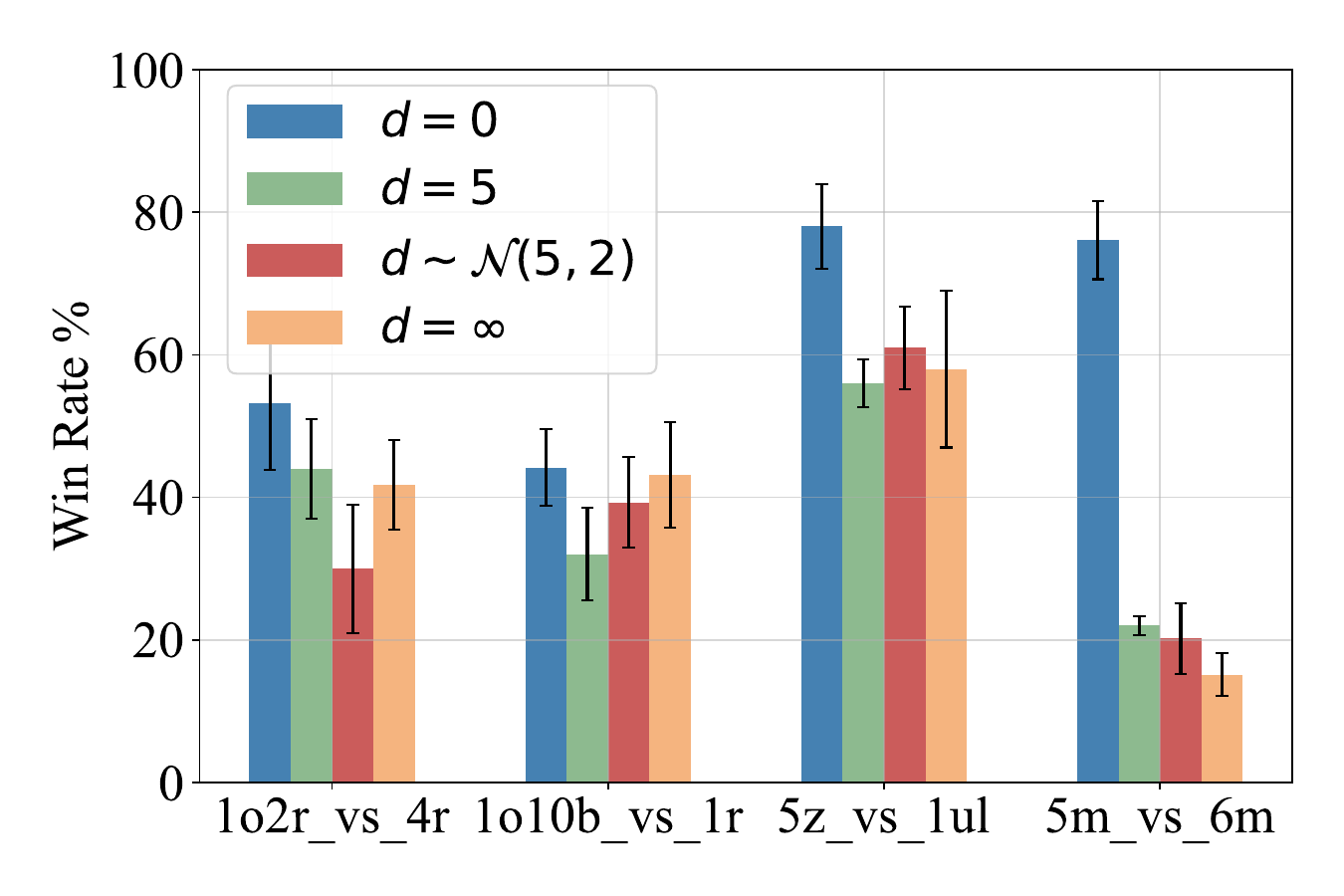} 
\caption{Performance of Communication-Enabled MARL Algorithm in Delayed Environments. Channel delays are configured as four types: 0, 5, infinity, and values sampled from a Gaussian distribution $\mathcal{N} \sim (5,2)$. Notably, the unit of delay is the decision time interval.}
\label{intro}
\end{figure}

We define the above issue of delays across decision intervals as {\it{Asynchronous Communication}}.
With delays, message transmission among agents lags, forcing them to rely on outdated messages to understand the environment and teammates, which may have already changed.
To verify this concern, we test conventional communication-enable MARL algorithms in SMAC \cite{samvelyan2019starcraft}, setting 4 end-to-end channel delays. As depicted in Fig. \ref{intro}, delays across decision intervals can impair the collaborative capability of original algorithms. Asynchronous messages can cause significant misguidance, potentially resulting in performance worse than without communication. This insight suggests that it is critical to research the robustness against communication delays across decision intervals. Here, we first summarize two types of communication delays in MARL:
\begin{itemize}
    \item \textbf{Fixed-Delay Setting}: Point-to-point channels encounter a fixed delay that exceeds the decision interval. Messages received are outdated but sent simultaneously.
    \item \textbf{Time-Varying-Delay Setting}: Point-to-point delay fluctuates over time and surpasses the decision interval. Messages received are sent at different times.
\end{itemize}

To achieve robustness against the above delays, we propose a novel communication MARL framework named CoDe.
Firstly, in terms of message expressiveness, we conceptualize intents as agents' future behavioral trends. 
Within a seq-to-seq structure, we design two regularization losses, the inference loss and the continuity loss, to extract and refine intents. 
The former enhances the representative capacity of behavioral trends by inferring future action sequences, while the latter reinforces stability over time by measuring intent similarity across adjacent time intervals.
Unlike past or present information, which can become diluted over time, delayed intents can still assist the recipient in understanding the situation.
Subsequently, we design a dual alignment mechanism to integrate multi-source asynchronous messages. 
The first alignment prioritizes messages related to their intents via an attention module, and the second one focuses on new messages based on the timeliness of transmission.
We implement the two types of delay settings in commonly used MARL benchmarks such as SMAC \cite{samvelyan2019starcraft}, GRF \cite{kurach2020google}, and Hallway \cite{Wang*2020Learning}. Extensive experiment results show that CoDe exhibits significant delay robustness.

Our contributions can be outlined as follows:
\begin{itemize}
    \item According to our knowledge, we are the first to investigate asynchronous communication in MARL and propose two delay settings.
    \item We introduce an intent extraction framework based on future action inference, which can represent agents' stable future behavioral trends.
    \item We design a dual alignment mechanism of intent and timeliness to integrate multi-source asynchronous messages, mitigating the impact of delays.
    \item We realize two delay settings in SMAC, GRF, and Hallway. CoDe consistently outperforms baseline under zero-delay, fixed-delay, and time-varying-delays.
\end{itemize}

\section{Related Works}
\subsubsection{Communication-Constrained MARL.}

Considering the constraints of limited channel bandwidth, several works concentrated on the sparsity of multi-agent communication \cite{foerster2016learning, zhang2019efficient, Wang*2020Learning, zhang2020succinct, yuan2022multi, sun2024t2mac, 10.5555/3635637.3663268}. NDQ \cite{Wang*2020Learning} designed two information-theoretic regularizers that maximize the mutual information between action selection and messages while minimizing the entropy of inter-agent messages. TMC \cite{zhang2020succinct} employed a time-window technique to mitigate the temporal variability of messages, thereby reducing the communication frequency. MAIC \cite{yuan2022multi} introduced an entropy regularization term to increase the concentration of attention and pruned communication links with low scores. T2MAC \cite{sun2024t2mac} quantified the necessity of communication by leveraging uncertainty reduction through different evidence ablation. Furthermore, some works \cite{mitchell2020gaussian,xue2021mis,yuan2024communication, ding2024learning, yu2024robust} focused on noisy or adversarial communication settings. To our best knowledge, there is a lack of research on the channel delay issue in multi-agent scenarios, which has motivated our study.

\subsubsection{Delay in RL.}
In the Markov Decision Process (MDP), delays are typically unavoidable, such as delays in state perception, action execution, and reward feedback. DATS \cite{chen2021delay} incorporated multi-step action delays into the environment model, effectively mitigating the impact of delays. DA-MAVL \cite{zhang2023multi} introduced a methodology for learning coarse-correlated equilibria under time-varying delays in reward feedback. DAMARL \cite{chen2020delay} researched observation delays in multi-agent settings by incorporating future actions into local observations to facilitate decision-making. DACOM \cite{yuan2023dacom} first integrated communication delays into multi-agent reinforcement learning, primarily focusing on decision-making of waiting time. Additionally, DACOM impractically assumed the delays within the interval between two consecutive decisions. In contrast, our research investigates explicitly the more general settings where channel delays surpass the decision interval, leading to asynchronous communication.

\begin{figure*}[t]
\centering
\includegraphics[width=1.0\linewidth]{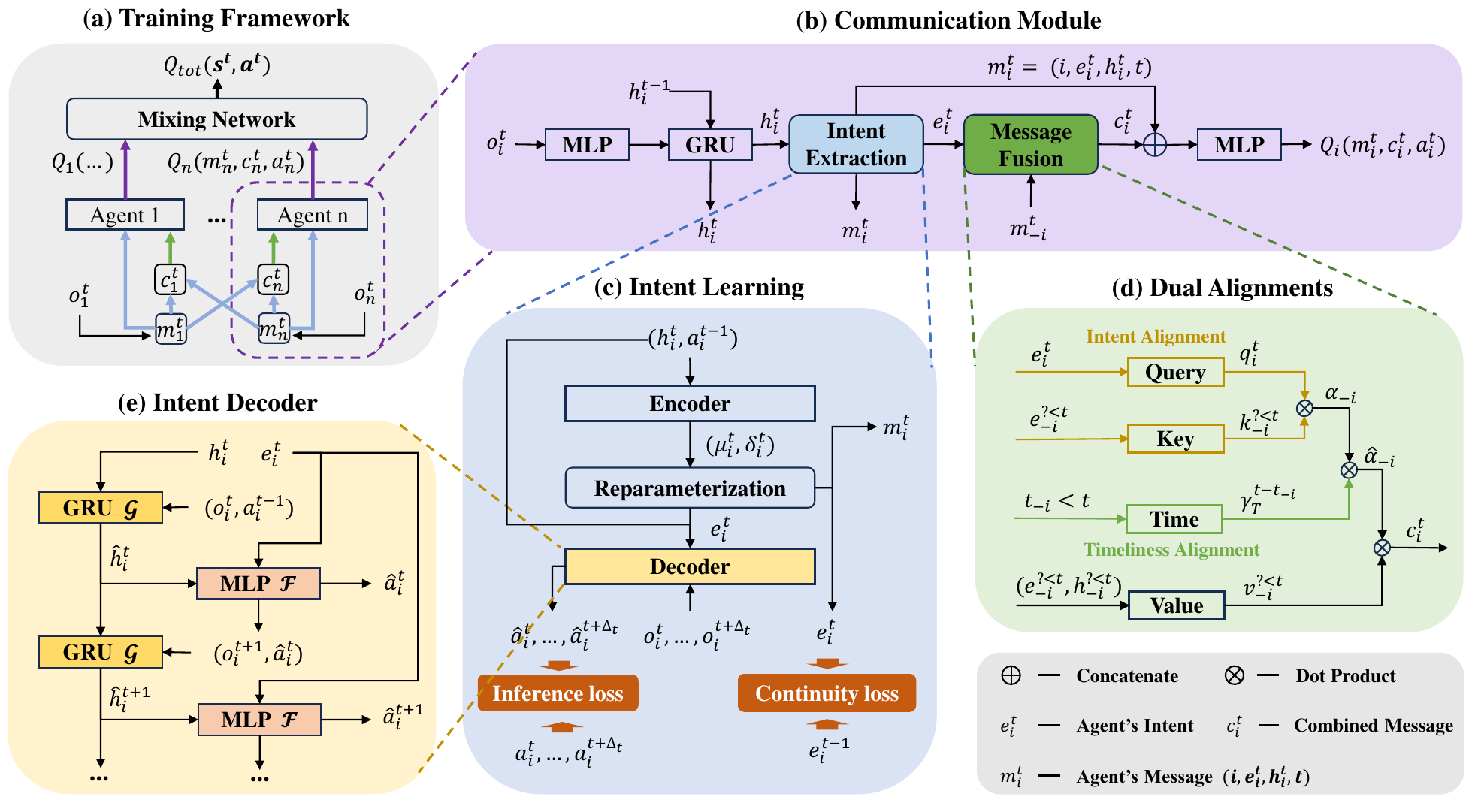} 
\caption{Overall Framework of CoDe. (a) The training framework; (b) The communication module consisting of intent extraction, message propagation, and message fusion; (c) The intent learning module via two designed losses; (d) The dual alignment module via a modified attention structure, in which $``-i"$ identifies the index of agents other than $i$ and $``?<t"$ represents certain timestamp before $t$; (e) The sequence prediction model used to decode the intent to future actions. }
\label{framework}
\end{figure*}

\section{Background}
Considering the problem of asynchronous communication, we first formulate multi-agent decision-making as Delay-Tolerant Decentralized Partially Observable Markov Decision Process (DT-Dec-POMDP), which can be formed as $\left \langle \mathcal{N}, \mathcal{S}, \Omega, \mathcal{O}, \mathcal{A}, \mathcal{R},\mathcal{P}, \gamma, \mathcal{M}, \mathcal{D} \right \rangle$. Specifically, $\mathcal{N}$ represents multiple agents in the system. $\mathcal{S}$ denotes the global state space that reflects an exhaustive environment overview. $\mathcal{O}$ signifies the partial observation space for agents and $\Omega$ represents the observation function that maps states to observations. $A$ refers to the joint action space. $\mathcal{R}$ stands for the reward function used to assess the quality of decisions. $\mathcal{P}$ acts as the environmental dynamics model, illustrating the influence of joint actions on states. $\gamma$ is the discount factor that balances long-term and short-term returns. Expanding the original Dec-POMDP, $\mathcal{M}$ delineates the message-passing within the system, which is characterized by point-to-point communication, facilitating the direct exchange of information between individual agents. $\mathcal{D} \in \{ \mathcal{D}_{f}, \mathcal{D}_{v} \}$ represents the instantaneous delay on the point-to-point channel, indicating that the messages received by agents at each moment were sent in the previous few moments.

\begin{itemize}
    \item \textbf{Fixed Delay $\mathcal{D}_{f}$}: $\forall_{i, j, t} d_{i,j}^t = d_f$, where ${d_{i,j}^t}$ denotes the delay between agent $i$ and $j$ at time $t$. $d_f$ is a fixed value;
    \item \textbf{Time-Varying Delay $\mathcal{D}_{v}$}: $d_{i,j}^t \sim \mathcal{N}(\mu, \delta)$, where $\mathcal{N}$ represents the predefined Gaussian distribution of delay.
\end{itemize}

\section{Method}
Our intuitive perspective suggests that in a delayed environment, messages should reflect the sender's stable behavioral trends while the receiver selectively utilizes outdated messages. Even if messages are delayed, receivers should still be able to extract some insights about the sender.

CoDe contains the mixing network in Fig. \ref{framework} (a) and the shared individual communication module in Fig. \ref{framework} (b). 
The core of CoDe is twofold: one aspect focuses on optimizing message design at the sender's end to extract and convey long-term intents (Fig. \ref{framework} (c, e)); the other aspect centers on optimizing message fusion at the receiver's end, employing a dual alignment of intent and timeliness to integrate multi-source asynchronous messages effectively (Fig. \ref{framework} (d)).
 
\subsection{Intent Learning Via Future Action Inference}
CoDe aims to learn an intent representation that can reflect agents' future behavioral trends to enhance inter-agent communication. Learned intents improve the collaborative abilities of agents in two aspects: firstly, transmitting intents aids in the comprehension of behaviors between agents, and secondly, intents can then act as verification for message fusion, enabling the selective use of multi-source messages.

Under channel delays, intents should possess two fundamental properties: (1) expressiveness of future behavior and (2) stability within a time window. Here, two regularization losses are devised to reinforce these two properties.
\subsubsection{Inference Loss.} 
To begin with, the inference loss is designed to augment the capacity to represent future action sequences. We believe that the historical trajectory of agents contains their behavioral motivations, which support agents in making coherent decisions. Hence, CoDe relies on the past to envision the future via an approximate seq-to-seq structure, depicted in Fig. \ref{framework} (c). 

To handle the historical trajectory $\tau = \left\{o_i^1, o_i^2,\ldots,o_i^t \right\}$, CoDe utilizes a GRU structure to obtain a compressed representation $h_i^t$. Subsequently, the preceding action $a_i^{t-1}$ is concatenated and fed into the encoder to obtain the distribution of intents $\mathcal{N}(\mu_i^t,\delta_i^t)$. Following reparameterization, we obtain the intent vector $e_i^t$ that preserves gradients. 

Afterward, a sequence prediction model is introduced to forecast actions for the next $K$ steps, as shown in Fig. \ref{framework} (e). Specifically, CoDe employs a GRU structure $\mathcal{G}$ and an action prediction network $\mathcal{F}$ for intent decoding rather than the commonly used transformer structure due to the limited sequence length. Firstly, we initialize the hidden state of $\mathcal{G}$ with the historical trajectory encoding $h^{t}$ at the current time step to enrich agents' comprehension of the past. In the following $K$ steps, the agent's current observations $o^{t+k}$ and the predicted action $\hat{a}^{t+k-1}$ from the previous time step are employed as inputs to $\mathcal{G}$. Then, the hidden state $\hat{h}^{t+k}$ processed through $\mathcal{G}$ and the intent $e^{t}$ from the initial time are passed to the prediction network $\mathcal{F}$ to forecast action. The difference between predicted actions $\hat{a}^{t+k}$ and ground-truth actions ${a}^{t+k}$ serves as the inference loss to train the network and refine the intent representation, defined as Eq. 1. In the case of discrete action space, $\| \cdot, \cdot\|$ denotes the cross-entropy loss. In contrast, for a continuous space, it signifies the mean squared error loss.
\begin{equation}
\begin{aligned}
    \mathcal{L}_{inf} = & \mathbb{E}_{\mathbf{\tau} \sim \mathcal{B}}\left[ \sum_{k=1}^{K-1} \| \mathcal{F}(e_i^t, \mathcal{G}( o_i^{t+k}, \hat{a}_i^{t+k-1})) , a_i^{t+k} \| \right. \\
    & \left. + \| \mathcal{F}(e_i^t, \mathcal{G}( o_i^{t}, a_i^{t-1})) , a_i^{t} \| \right].
\end{aligned}
\end{equation}


\subsubsection{Continuity Loss.}
Given that intent reflects future behavioral trends, it is expected to exhibit continuity over the short term. We compute the cosine similarity between intents generated at adjacent time steps and maximize it to enhance the consistency of intents, formulated as: 
\begin{equation}
    \mathcal{L}_c = \mathbb{E}_{\tau \sim \mathcal{B}} \left[- \frac{e_i^{t-1} \cdot e_i^t}{\|e_i^{t-1}\|\|e_i^t\|} \right].
\end{equation}

Demanding stable intents via Eq. 2 can narrow the intent distribution, limiting exploration. Hence, we introduce an auxiliary loss $\mathcal{L}_k$ to increase the diversity of intents. Formally, we compute the Kullback-Leibler divergence between the intent distribution and a standard normal distribution $D_{KL}(\mathcal{N}(\mu_i^t,\delta_i^t) \|  \mathcal{I})$. $\mathcal{L}_k$ can be formulated as Eq. 3, in which $n$ denotes the dimension of intents. 
\begin{equation}
    \mathcal{L}_{k} = \mathbb{E}_{\tau \sim \mathcal{B}} \left[-\frac{1}{2} \sum_{i=1}^{n} \left( \delta_{i}^2 + \mu_{i}^2 - 1 - \log(\delta_{i}^2) \right) \right].
\end{equation}
In addition to the training loss for subsequent reinforcement learning, we design two regularization losses and an auxiliary loss to optimize intent representation, balancing their effects through hyperparameters formed as:
\begin{equation}
    \mathcal{L}_{int} = \lambda_{inf}\mathcal{L}_{inf} + \lambda_c\mathcal{L}_c + \lambda_k\mathcal{L}_k.
\end{equation}

\subsection{Dual Alignment Of Intent and Timeliness}
We posit that, even though the sender can promptly convey its intents, the receiver encounters challenges in managing asynchronous messages from multiple sources. 
In the CoDe framework, the structure of a message is represented as $\left \langle i, e_i^t, h_i^t, t \right \rangle$. $e_i^t$ represents the intent involving the agent's future behavioral trend, $h_i^t$ signifies the message content encompassing the agent's insights, and $t$ denotes the delivery time.
To prioritize the integration of messages, we propose a dual alignment mechanism in Fig. \ref{framework} (d). On the one hand, the learned intent can serve as credentials, prompting agents with similar intents to pay more attention to each other. On the other hand, delay also acts as a criterion, with newer messages carrying greater significance for the recipient.

\subsubsection{Intent Alignment.} Leveraging intent as the credential, we utilize an attention mechanism \cite{vaswani2017attention} for the initial alignment. The attention mechanism is extensively employed in multi-source information fusion for its capacity to extract features regardless of sequence. In our design, the receiver's intent $e_i$ acts as the query, the sender's intent $e_{-i}$ serves as the key, and the sender's content combined with intents $(e_{-i}, h_{-i})$ serves as the value. Formally, the attention score is calculated as follows:
\begin{equation}
    \alpha_{i,j}=\frac{\exp\left(\frac1{\sqrt{d_k}}\cdot{e_i}W^Q\cdot\left(e_jW^K\right)^\top\right)}{\sum_{u=1}^{-i}\exp\left(\frac1{\sqrt{d_k}}\cdot{e_i}W^Q\cdot\left(e_uW^K\right)^\top\right)},
\end{equation}
where $W^Q$ and $W^K$ represent parameter matrices for linearly transforming queries and keys. $\frac{1}{\sqrt{d_k}}$ serves as a scaling factor for dot-product attention. $\alpha_{i,j}$ defines the correlation between agent $j$'s message and agent $i$. To enhance attention concentration, we drew inspiration from \cite{yuan2022multi}, introducing an entropy regularization term to minimize uncertainty by reducing entropy. 
\begin{equation}
\mathcal{L}_e=-\lambda_e\sum_{i=1}^n\mathcal{H}(\alpha_i.)=-\lambda_e\sum_{i\neq j}^n\alpha_{ij}\log\alpha_{ij}.
\end{equation}

\subsubsection{Timeliness Alignment.} Due to delays, the time lag between sending and receiving an intent can be substantial. In such instances, the sender's intent may have evolved, and aligning with outdated intents could easily lead to the breakdown of collaborative behaviors. 
Here, we employ a straightforward yet effective approach to address asynchronous messages by a secondary decay. In detail, CoDe incorporates a temporal discount factor $\gamma_T$ for exponential decay to the weight of intent alignment $\alpha_{i,j}$. As shown in Eq. 6, $\Delta t$ represents the time interval since the message $m_j$ was sent.
\begin{equation}
    \hat{\alpha}_{i,j} = \alpha_{i,j} * {\gamma_T}^{\Delta t}.
\end{equation}

Following the dual alignment process, we derive attention weights that balance intent and timeliness. Subsequently, weighted message fusion is carried out, expressed as 
\begin{equation}
    c_i^t = \sum_{j=1}^{-i} \hat{\alpha}_{i,j} (e_j,h_j) W^V, 
\end{equation}
where $W^V$ denotes the linear transformation for part of messages $(e_j,h_j)$. Notably, we assume zero channel delays during training. Hence, Eq. 5, 7, and 8 vary between the training and testing phases.

\begin{figure*}[ht]
\centering
\includegraphics[width=1.0\linewidth]{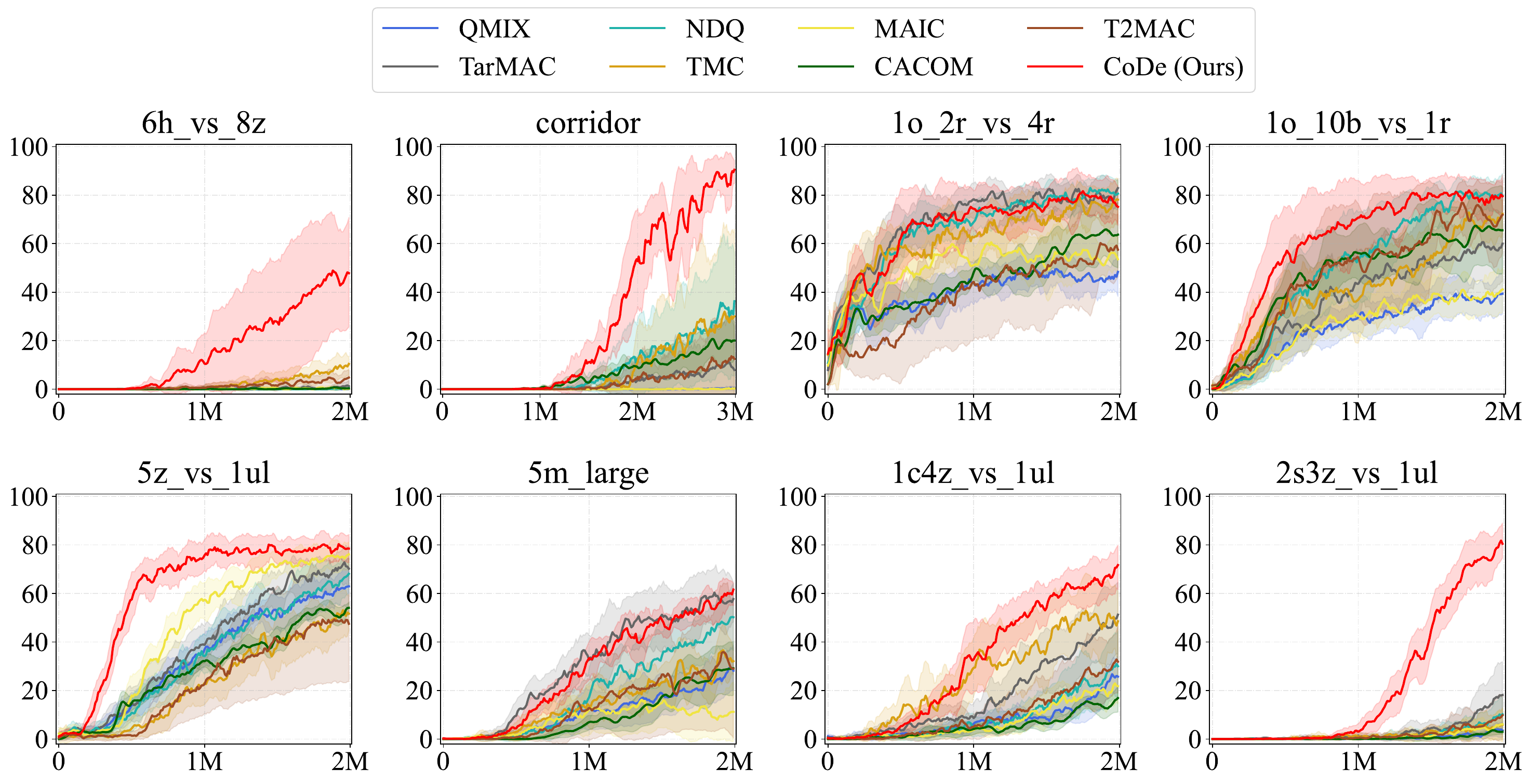} 
\caption{Algorithms Performance in SMAC under Zero Communication Delay. Each curve represents the average result of 5 random seeds. The last three are our proposed new maps.}
\label{SMAC_RESULTS}
\end{figure*}

\subsection{Overall Training Objective}
After message fusion, the integrated information is incorporated into the MARL training process. This paper adopts the foundational framework QMIX \cite{rashid2020monotonic} while also adaptable to any value decomposition algorithm like VDN \cite{sunehag2017value} or QPLEX \cite{wang2020qplex}.
$\mathcal{L}_{RL}$ is defined as the standard temporal difference (TD) loss using $Q_{tot}$ from the mixing network. By integrating intent learning and the dual alignment mechanism, the complete training loss is as follows:
\begin{equation}
    \mathcal{L}_{tot} = \mathcal{L}_{RL} + \mathcal{L}_{int} + \mathcal{L}_{e}.
\end{equation}

\subsection{Delayed Communication Protocol}
After training, we comprehensively describe the decision-making process in delayed scenarios, as illustrated in Fig. \ref{framework} (b). At each time step, the agent perceives the local environment to gather partial observations $o_i^t$, which are then encoded into a historical trajectory representation $h_i^t$. The current intent distribution is derived by feeding  $h_i^t$ and the previous action $a_i^{t-1}$ into the intent encoder, and the intent $e_i^t$ is obtained after reparameterization sampling. Then, the agent constructs and broadcasts messages $m_i^t$. Concurrently, as a recipient, the agent maintains a message buffer that stores the most recent messages from other agents and performs dual alignments to integrate asynchronous messages $m_{-i}^{?<t}$ from diverse sources. The integrated information $c_i^t$, alongside its own messages $m_i^t$, is jointly fed into the local Q-values $Q_i(\cdots)$ for decision-making. The pseudocode for training and testing can be found in the Appendix.

\section{Experiments}
To evaluate CoDe, we conduct experiments on three benchmarks, including Hallway \cite{Wang*2020Learning}, SMAC \cite{samvelyan2019starcraft}, and GRF \cite{kurach2020google}. 
In the section, we select seven algorithms as baselines, including QMIX \cite{rashid2020monotonic}, TarMAC \cite{das2019tarmac}, NDQ \cite{Wang*2020Learning}, TMC \cite{zhang2020succinct}, MAIC \cite{yuan2022multi}, CACOM \cite{li2023context} and T2MAC \cite{sun2024t2mac}.
 We aim to validate the following three questions experimentally:
(1) Does the intent-based communication mechanism exhibit superior performance in a zero-delay environment?
(2) Does our proposed CoDe demonstrate robustness under various communication delay conditions?
(3) Does the learned intent effectively represent the future action trend of agents?

\begin{table*}[t]
\centering
\fontsize{9}{11}\selectfont  
\label{tab:fixeddelay}
\begin{tabularx}{\textwidth}{p{0.8cm}*{7}{>{\centering\arraybackslash}X}}
\toprule
\centering Delay & Map & 1o\_2r\_vs\_4r & 1o\_10b\_vs\_1r & 5z\_vs\_1ul & 5m\_large & 1c4z\_vs\_1ul & 2s3z\_vs\_1ul \\
\midrule
\centering \multirow{5}{*}{$3$} & NDQ & $\mathbf{63.1 \pm 7.4}$ & $52 \pm 3.1$ & $39.2 \pm 4.5$ & $\mathbf{39.6 \pm 3.1}$ & $6.3 \pm 1.4$ & $8.2 \pm 1.7$ \\
&TarMAC & $33.3 \pm 4.5$ & $47.2 \pm 5.2$ & $48.4 \pm 7.9$ & $28.2 \pm 6$ & \underline{$61.1 \pm 10.5$} & \underline{$43.1 \pm 6.9$} \\
&MAIC & $48 \pm 7.5$ & $26.2 \pm 8.4$ & \underline{$66.8 \pm 5.8$} & $8.4 \pm 3.4$ & $14.4 \pm 6.4$ & $6 \pm 3.6$ \\
&T2MAC & $56.3 \pm 3.3$ & \underline{$55.3 \pm 9.1$} & $43.8 \pm 7.6$ & $13.1 \pm 5.2$ & $15.3 \pm 3.7$ & $15.1 \pm 4.4$ \\
&CoDe & \underline{$60.8 \pm 7.5$} & $\mathbf{59.3 \pm 8.1}$ & $\mathbf{67.7 \pm 4.3}$ & \underline{$33.7 \pm 1.4$} & $\mathbf{72 \pm 5.8}$ & $\mathbf{89.3 \pm 3.3}$ \\
\midrule
\centering \multirow{5}{*}{$5$}&  NDQ & $48.2 \pm 3.9$ &\underline{ $46.7 \pm 4.5$} & $37.8 \pm 3.2$ & \underline{$25.4 \pm 4.1$} & $5.5 \pm 1.7$ & $7.7 \pm 0.8$ \\
&TarMAC & $24.7 \pm 4.8$ & $39.5 \pm 6.5$ & $42.7 \pm 9.6$ & $12 \pm 3.7$ & \underline{$56.4 \pm 5.9$} & \underline{$32.8 \pm 5.4$} \\
&MAIC & $44.6 \pm 5.7$ & $25.1 \pm 6.1$ & \underline{$61.3 \pm 7.2$} & $6.2 \pm 4.1$ & $11.1 \pm 5.4$ & $4.6 \pm 2.8$ \\
&T2MAC & \underline{$55.1 \pm 6.4$} & $43.5 \pm 8.8$ & $39.3 \pm 6.7$ & $3.5 \pm 2.4$ & $13.7 \pm 4.5$ & $17.1 \pm 5.8$ \\
&CoDe & $\mathbf{52.3 \pm 6.1}$ & $\mathbf{49.8 \pm 9.3}$ & $\mathbf{68.6 \pm 7.5}$ & $\mathbf{29.7 \pm 2.3}$ & $\mathbf{64 \pm 6.1}$ & $\mathbf{87.3 \pm 4.9}$ \\
\bottomrule
\end{tabularx}
\caption{Results in SMAC under Fixed Communication Delay. Bold and underlined represent the first and second performances.}
\end{table*}

\begin{table*}[t]
\centering
\fontsize{9}{11}\selectfont  
\setlength{\tabcolsep}{1mm}
\begin{tabularx}{\textwidth}{p{1.2cm}*{7}{>{\centering\arraybackslash}X}}
\toprule
\centering Delay & Map & 1o\_2r\_vs\_4r & 1o\_10b\_vs\_1r & 5z\_vs\_1ul & 5m\_large & 1c4z\_vs\_1ul & 2s3z\_vs\_1ul \\
\midrule
\centering \multirow{5}{*}{$\mathcal{N}(3,2)$}&NDQ & \underline{$66 \pm 7.8$} & \underline{$59.2 \pm 4$} & $41.3 \pm 6$ & \underline{$34.8 \pm 3.8$} & $6.4 \pm 1.3$ & $8.1 \pm 2.3$ \\
&TarMAC & $47.1 \pm 4.6$ & $45.2 \pm 5.4$ & $44.7 \pm 5.3$ & $10.5 \pm 1.6$ & \underline{$48.2 \pm 3.1$} & \underline{$20.9 \pm 2.6$} \\
&MAIC & $49.1 \pm 4.5$ & $27.6 \pm 4.4$ & \underline{$67.7 \pm 5.8$} & $16.9 \pm 3.8$ & $14.6 \pm 4.1$ & $6.8 \pm 1.6$ \\
&T2MAC & $59.5\pm 4.1$ & $55.1 \pm 6.1$ & $47.5 \pm 5.3$ & $15.4 \pm 3.4$ & $17.7 \pm 3.3$ & $18 \pm 2.5$ \\
&CoDe & $\mathbf{69.2 \pm 6.3}$ & $\mathbf{62.1 \pm 7.5}$ & $\mathbf{72.2 \pm 4.8}$ & $\mathbf{36.7 \pm 2.1}$ & $\mathbf{71 \pm 5.3}$ & $\mathbf{88 \pm 2.9}$ \\
\midrule
\centering \multirow{5}{*}{$\mathcal{N}(5,2)$}&NDQ & $50.3 \pm 3.1$ & \underline{$50.8 \pm 3.4$} & $38.8 \pm 6.1$ & $\mathbf{36.7 \pm 4.6}$ & $5.7 \pm 1.2$ & $7.6 \pm 1.1$ \\
&TarMAC & $38.1 \pm 6.9$ & $38.3 \pm 4.1$ & $38.5 \pm 5.5$ & $4.5 \pm 1.4$ & \underline{$49 \pm 7.8$} & \underline{$21.8 \pm 2.3$} \\
&MAIC & $44.9 \pm 5$ & $27.3 \pm 4.2$ & \underline{$59.6 \pm 4.1$} & $7.9 \pm 2.1$ & $14.9 \pm 4.1$ & $6.2 \pm 4.3$ \\
&T2MAC & \underline{$57.8 \pm 6$} & $48.1 \pm 7.2$ & $42.1 \pm 5$  & $6.4 \pm 7.6$ & $16.8 \pm 2.8$ & $14.9 \pm 2.6$ \\
&CoDe & $\mathbf{58.9 \pm 8.5}$ & $\mathbf{54.1 \pm 10.4}$ & $\mathbf{66.7 \pm 5}$ & \underline{$28.3 \pm 3.8$} & $\mathbf{67.4 \pm 4.6}$ & $\mathbf{88 \pm 2.7}$ \\
\bottomrule
\end{tabularx}
\label{unfixeddelay}
\caption{Results in SMAC under Time-Varying Communication Delay.}
\end{table*}

\subsection{Performance In Delay-free Environments}
We evaluate the algorithm under zero-delay conditions to assess the effectiveness of inter-agent intent transmission. Since CoDe does not consider communication concurrency, for fairness, the experiments in this section employ a fully connected communication structure for all other baselines.

\subsubsection{SMAC.}

\begin{figure}[t]
\centering
\includegraphics[width=1.0\linewidth]{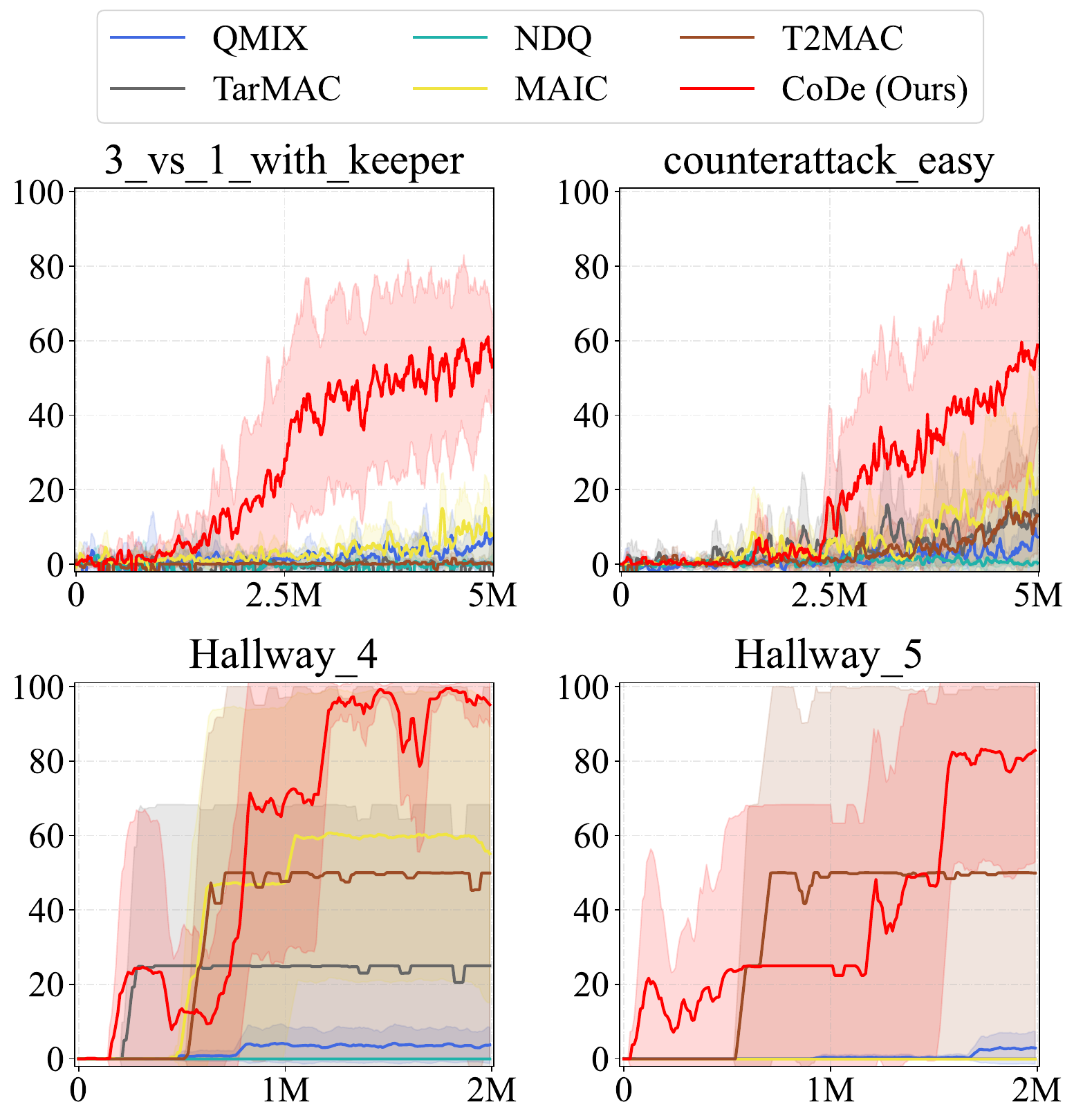} 
\caption{Performance in GRF and Hallway under Zero Communication Delay. The first line represents the result in GRF, while the second line represents the result in Hallway.}
\label{Hallway_football}
\end{figure}

As the most popular MARL benchmark, SMAC exhibits limited suitability under communication conditions due to constrained environmental randomness and local observability. Hence, apart from two original super-hard maps and three communication-enhanced maps introduced by NDQ, we additionally devise three supplementary maps, as shown in the Appendix. In our designed maps, the initial positions of the units are randomized, necessitating agent exploration and identification of both allies and enemies. Furthermore, to defeat enemies, agents must develop tactical strategies, underscoring the critical role of communication. As depicted in Fig. \ref{SMAC_RESULTS}, CoDe consistently demonstrates impressive performance and sampling efficiency across nearly all maps. Comparably, most baselines struggle to produce effective results on the newly introduced maps. In scenarios with high levels of randomness and the requirement for collaborative exploration, we believe that agents should transmit messages about themselves rather than suggestions to others. As for how the messages are utilized, it should be left to the receiving agents to determine.

\subsubsection{GRF.}
We further evaluate CoDe on the more challenging GRF; the results are depicted in Fig. \ref{Hallway_football}. Compared with baselines, CoDe achieves high team scoring rates across two tasks. Due to the more vital interactivity involving actions like passing among agents in GRF, conveying one's behavioral intents enhances the coordination of agents' actions.

\subsubsection{Hallway.}
In the Hallway benchmark, multiple agents navigate parallel linear tracks by moving left and right. The maximum reward is achieved when all agents simultaneously reach the goal for the first time. We devise two tasks to control the coordinated movement of 4 and 5 agents. As illustrated in Fig. \ref{Hallway_football}, CoDe achieves remarkable performance across two tasks compared to the baseline. Building upon pervasive communication, we attribute this performance to future behavioral intents within CoDe. In high-risk environments such as Hallway, incorporating long-term intents proves advantageous for coordinating the agents' actions and mitigating risky decisions. In contrast, prior works like NDQ and TarMAC tend to convey behavioral suggestions to other agents, which leads to a relatively unclear comprehension of the sender's state and intents for the recipients.

\subsection{Performance In Delayed Environments}
Next, we implement two communication delay settings to assess the algorithm's robustness, involving sampling point-to-point channel delays from either a Gaussian distribution or a fixed value at each time step. Notably, to handle 
out-of-time-order messages, we discard any older timestamped messages received after the arrival of a message with a new timestamp. Additionally, considering the need for extra storage space in CoDe, we augment the storage capacity of other baseline algorithms for fairness. In cases where new messages are unavailable, agents can rely on the most recent messages to make decisions.

\subsubsection{Results.} We perform two sets of experiments in SMAC, one with fixed delays and the other with varying ones. In Tab. 1 and 2, we observe that CoDe maintains stable performance across various delay settings, validating the effectiveness of the dual alignment mechanism. Under fixed delays, the performance of CoDe slightly declines compared to varying delays. This can be attributed to the synchronized message sources, which weaken timeliness alignment. The consistent timestamps of messages lead to uniform down-weighting during information fusion, causing agents to focus more on themselves and reducing collaboration. Notably, prior works, such as NDQ and TarMAC, which center on communication sparsity, also exhibit a degree of delay robustness. The sparse communication structure implies that only a small subset of messages influences decision-making, alleviating the negative impact of delays. 


\begin{table}[t]
\centering
\fontsize{9}{11}\selectfont  
\setlength{\tabcolsep}{1mm}
\label{tab:ablationalignment}
\begin{tabularx}{\linewidth}{p{1.8cm}p{0.9cm}*{3}{>{\centering\arraybackslash}X}}
\toprule
\centering Maps & CoDe &CoDe \newline w/o IA &CoDe \newline w/o TA & CoDe \newline w/o DA \\
\midrule
\centering 1o\_10b\_vs\_1r &\centering $\mathbf{62.1}$ & $35.6$ & $58.5$ & $33.8$ \\
\centering 1c4z\_vs\_1ul &\centering $\mathbf{71}$ & $60.4$ & $65.8$ & $59.5$\\
\centering 2s3z\_vs\_1ul &\centering $\mathbf{88}$ & $79.6$ & $83.7$ & $78.3$\\
\bottomrule
\end{tabularx}
\caption{Ablation over alignment modules. ``IA" means the intent alignment, ``TA" means the timeliness alignment, and ``DA" means dual alignments.}
\end{table}

\subsubsection{Ablation.} To validate the dual alignment mechanism, we execute an ablation experiment in Tab. 3. Results show that removing any of the alignment modules leads to a decline in the delay robustness of CoDe, demonstrating the effectiveness of dual alignment. Comparably, intent alignment exhibits a more significant impact.

\subsection{Analysis Over The Learned Intent}
We analyze the learned intent by visualizing its capability to represent future behaviors while conducting ablation experiments on various losses during the intent training process.
\subsubsection{Visualization.}
\begin{figure}[t]
\centering
\includegraphics[width=1.0\linewidth]{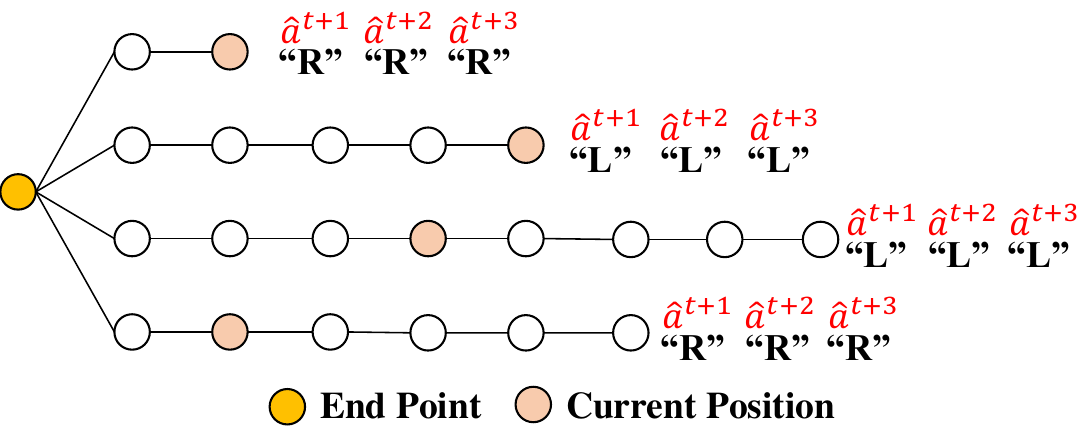} 
\caption{Visualization of the learned intent. 
"L" and "R" represent the left and right actions, respectively. $\hat{a}^{t+?}$ refers to the expected action at a future time, as decoded from the intent at time $t$.}
\label{hallway_vis}
\end{figure}

We decode the learned intents, converting them into future sequences of actions, and visualize them within the Hallway environment. From Fig. \ref{hallway_vis}, it can be observed that decoded action sequences reflect the accurate behavioral trends of agents at the current time step.

\subsubsection{Ablation.}

\begin{figure}[t]
\centering
\includegraphics[width=1.0\linewidth]{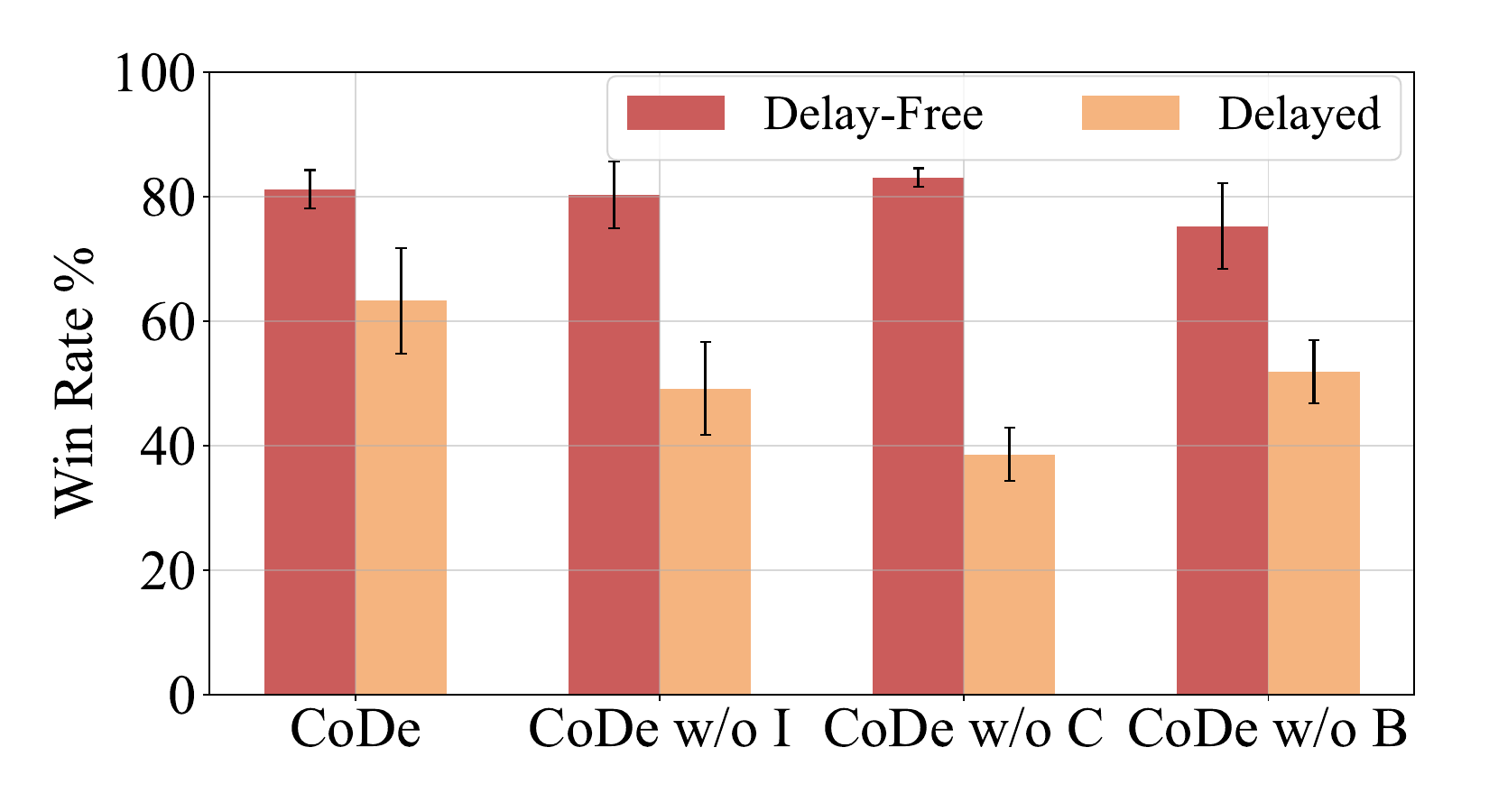} 
\caption{Ablation over two regularization losses. ``I" means the inference loss, ``C" means the continuity loss, and ``B" means both losses.}
\label{Ablation_loss}
\end{figure}

A set of ablation experiments is implemented to verify the effectiveness of the proposed inference loss and continuity loss. Fig. \ref{Ablation_loss} captures the performance changes of the ablated CoDe in both delay-free and delayed scenarios. Evidently, in the absence of delays, removing losses does not yield significant performance variations. This can be attributed to the immediate message transmission, which does not demand long-term expressive and stable intent representation. However, in the presence of delays, any form of ablation leads to a noticeable performance deterioration, confirming the effectiveness of the two losses. Regarding CoDe w/o B, the learned intent vectors tend to converge towards random variables upon eliminating two losses, diminishing their communicative efficacy. Consequently, the impact of delayed random intent is also relatively minor.

\section{Conclusions}
This study explores a new field concerning the asynchronous communication issue in MARL under channel delays. To address this issue, we propose a new communication framework called CoDe. For senders, CoDe extracts and conveys intents that reflect future behavioral trends. For receivers, CoDe introduces a dual alignment mechanism of intents and timeliness to merge asynchronous messages. Empirical findings indicate that CoDe outperforms baselines across various delay scenarios. However, CoDe is limited by the length of future action prediction and struggles with more severe delays, a challenge we aim to address in future work. We aspire that this work will inspire increased interest in the research of MARL within communication delay scenarios.

\section{Acknowledgments}
This work was supported by the National Natural Science
Foundation of China (Grant No. 62206013).
\bibliography{aaai25}

\end{document}